\documentclass[preprint]{elsarticle}

\usepackage{cprotect}
\usepackage{xcolor} 
\usepackage{dcolumn}
\usepackage{bm}
\usepackage{float}
\usepackage{setspace}
\usepackage{comment}
\usepackage{adjustbox}
\usepackage{placeins}
\usepackage{lineno}

\usepackage{upgreek}
\usepackage{makecell}
\usepackage{subcaption}

\usepackage{float}
\usepackage{caption,lipsum}

\usepackage[english]{babel}
\usepackage[utf8]{inputenc}
\usepackage{geometry}

\usepackage{amsmath}
\usepackage{amsfonts}
\usepackage{csquotes}
\usepackage{url}
\usepackage{hyperref}
\usepackage{physics}
\AtBeginDocument{\RenewCommandCopy\qty\SI}
\usepackage[separate-uncertainty=true, separate-uncertainty-units = single]{siunitx}
\usepackage[version=4]{mhchem}
\usepackage{subcaption}

\DeclareUnicodeCharacter{2212}{\ensuremath{-}}
\usepackage{etoolbox} 

\usepackage{graphicx}

\journal{Nuclear Instruments and Methods in Physics Research A}
\myfooter[L]{Preprint submitted to Nuclear Instruments and Methods in Physics Research A}
\begin{document}

\begin{frontmatter}
\title{Data-Driven Calibration of Large Liquid Detectors with Unsupervised Learning}

\author{Scott DeGraw\corref{cor1}}
\cortext[cor1]{Corresponding author}
\ead{scott.degraw@physics.ox.ac.uk}
\author{Steve Biller}
\author{Armin Reichold}
\address{University of Oxford, The Denys Wilkinson Building, Keble Road, Oxford, OX1 3RH}

\begin{abstract}
This paper demonstrates a novel method to extract photomultiplier tube (PMT) calibration timing constants in large liquid scintillation detectors from physics data using the machinery of unsupervised deep learning. The approach uses a simplified physical model of optical photon transport in the loss function, with PMT calibration constants treated as free parameters, and the simple assumption that individual events represent point-like emission. The problem is, thus, effectively reduced to that of regression on a very large scale, made tractable by deep learning architectures and automatic differentiation frameworks. Using data from the 9,300 PMTs in the SNO+ detector, the method has been shown to reliably extract 3 calibration constants for each of the over 7,500 online PMTs using radioactive background events. We believe that this basic approach can be straightforwardly generalized for a wide range of applications.
\end{abstract}


\end{frontmatter}

\section{Introduction}
\begin{figure}[htb]
    \begin{subfigure}[t]{0.45\columnwidth}
        \centering
        \includegraphics[width=1\columnwidth]{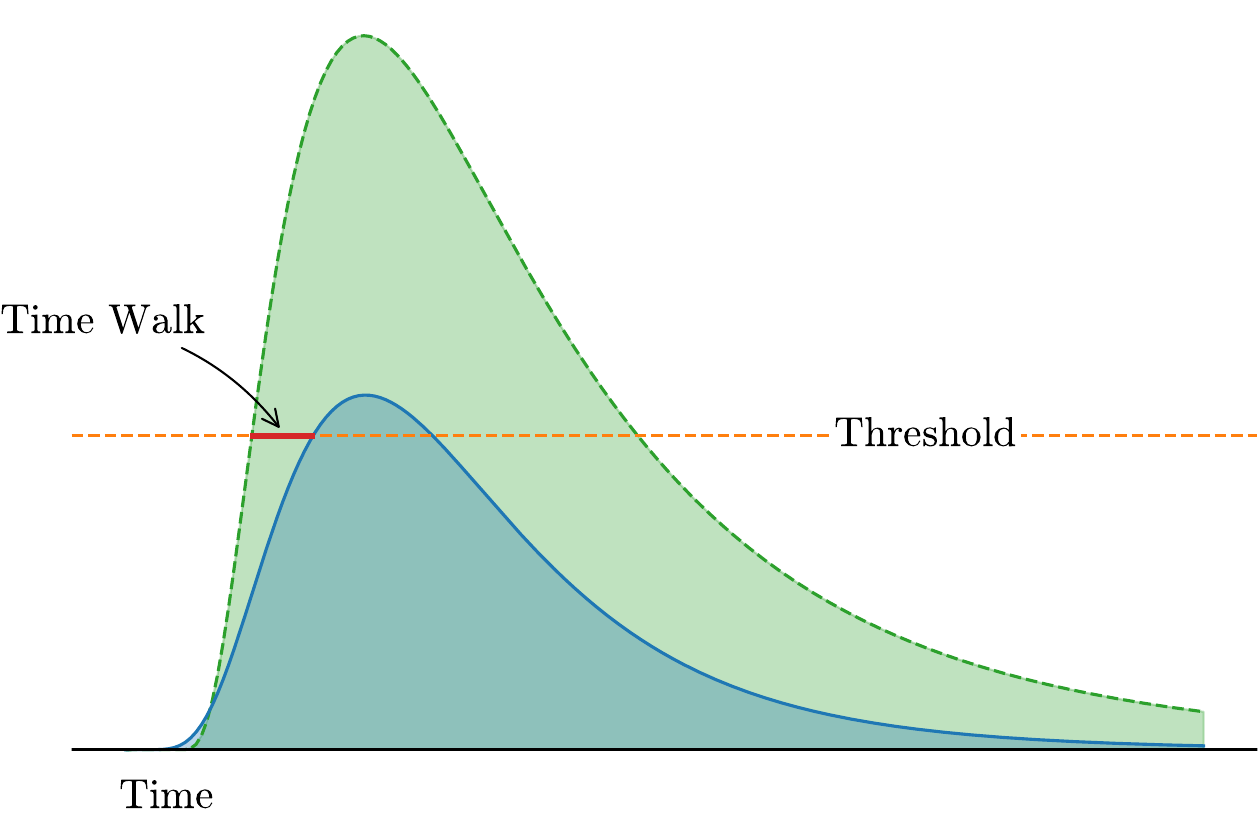}
        \caption{An illustration of the time walk effect. The dashed line indicates a threshold applied to a typical PMT signal.}
    \end{subfigure}%
    \hfill
    \begin{subfigure}[t]{0.45\columnwidth}
        \centering
        \includegraphics[width=1\columnwidth]{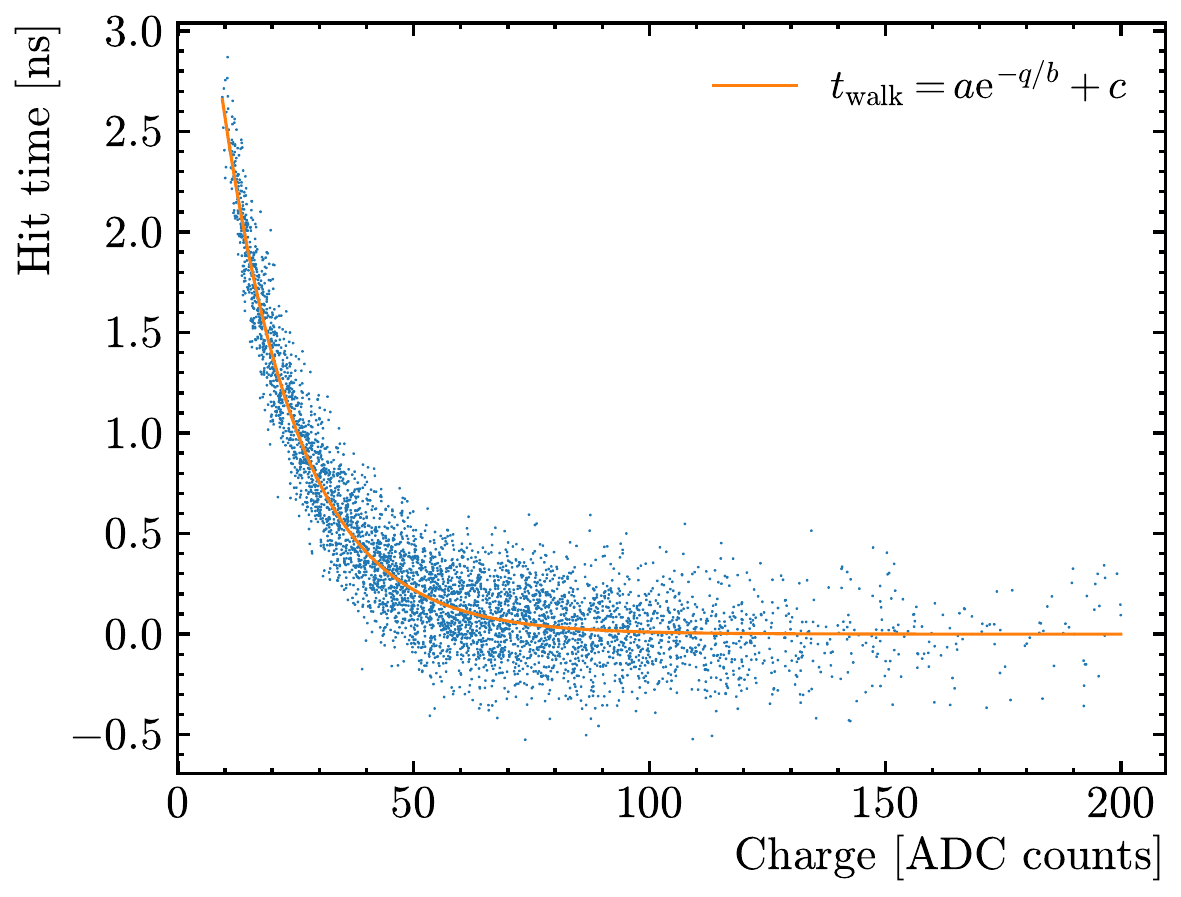}
        \caption{A simulated demonstration of a time walk. Values are typical. Hit times are measured relative to flat part of time walk.}
    \end{subfigure}
    \caption{Time walk effect}
    \label{fig:time_walk_illustration}
\end{figure}

Large liquid scintillator detectors involve thousands of photomultiplier tubes (PMTs) to detect the scintillation light produced by charged particles depositing energy in the scintillating medium.
The timing calibration of these PMTs (typically to the level of a few tenths of a nanosecond) is crucial to permit the accurate reconstruction of event positions based on the relative PMT hit times.
For this, multiple effects need to be taken into account.
Each PMT is connected to a data acquisition system via cables and front-end electronics that result in different time delays.
To reduce noise hits, PMTs are only triggered when charge signals cross some pre-defined threshold.
This threshold gives rise to a charge-dependent \textit{time walk} effect where higher charge pulses trigger the PMT earlier leading to earlier hit times. 
The precise character of the time walk can vary across PMT channels.
An illustration of this effect is shown in \autoref{fig:time_walk_illustration}.

Standard methods for performing this type of calibration involve dedicated calibration light sources, either deployed \textit{in situ} or fixed inside the detector.
Deployed systems, such as laserballs (also known as diffuser balls) used in SNO \cite{moffatOpticalCalibrationHardware2005}, SNO+ \cite{valderLaserballCalibrationDevice2024}, Super-Kamiokande \cite{super-kamiokandecollaborationCalibrationSuperKamiokandeDetector2014} and JUNO \cite{zhangLaserCalibrationSystem2019} deploy a stationary light source and pulse light at known positions and times.
These involve dedicated campaigns that are limited in how frequently they can be deployed, involve a significant amount of manual work, use light with different characteristics than physics events, generally stop physics data taking, and can carry a risk of contaminating the detector.
Fixed systems in SNO+ \cite{falkCommissioningELLIESNO2017}, Borexino \cite{caccianigaMultiplexedOpticalfiberSystem2003} and planned usage in Hyper-Kamiokande \cite{vinningNarrowbeamDiffuserSubsystem2019}, typically use sources that are not in the detection volume and similarly produce light with wavelength spectra and timing that are different from those of physics events and, hence, can involve some extrapolation.
An in situ method that uses data from actual events, without the requirement of specialised hardware, would therefore carry a number of advantages, provided that the calibration constants could be reliably extracted over a relatively short length of time.

The approach demonstrated here involves known radioactive background events in the SNO+ detector collected during regular physics data taking to perform a regression fit to 3 calibration parameters per PMT that describe individual time offsets and charge-dependent walk corrections for a total of over 22,000 parameters. This is made easily tractable through the use of the robust gradient descent methods employed in automatic differentiation frameworks which are well suited to this task.

\section{Data-Driven Calibration}
\subsection{Detector}

The target volume of SNO+ is a \qty{12}{\metre} diameter acrylic vessel (AV) filled with \qty{780}{\tonne} of liquid scintillator.
The AV is surrounded by an \qty{18}{\metre} diameter PMT support structure holding approximately 9,300 inward facing PMTs, around 7,500 of which were online and had well calibrated electronics in the dataset used.
The AV and support structure both reside \qty{2}{\kilo\metre} underground in a cavity excavated in rock filled with \qty{7}{\kilo\tonne} of ultra pure water. 
The AV is offset in the vertical direction to the centre of the PMT support structure by approximately \qty{18.5}{\centi\meter} and has a neck for filling at the top.
The PMTs are Hamamatsu R1408 PMTs inherited from SNO.
The liquid scintillator consists of Linear Alkyl Benzene (LAB) as a solvent and \qty{2.2}{\g\per\litre} of 2,5-diphenyloxazole (PPO) as a secondary fluor.
The current phase of SNO+ as of December 2023 additionally contains a bisMSB wavelength shifter to improve light collection, but the dataset used here was taken before its addition.
Comprehensive descriptions of the SNO, and SNO+ detectors can be found in \cite{thesnocollaborationSudburyNeutrinoObservatory2000,thesno+collaborationCurrentStatusFuture2016}, respectively.

\subsection{Timing Model}

To sufficiently model the timing characteristics for PMT $i$, a decaying exponential as a function of PMT charge, $q$, was used to parametrize the time walk correction, in addition to a constant delay, $c_i$:

\begin{equation}\label{eq:time_walk_model}
    \tau_i(q) = a_i \exp(-q / b_i) + c_i.
\end{equation}

The charge metric used is the PMT signal integrated over \qty{70}{\nano\second} measured in 12 bit analogue-to-digital counts.
Adding an additional linear term with charge was tried but found to be unnecessary.

A priori, the time walk should be a monotonically decreasing function with charge $q$, meaning $a_i \ge 0$ and $b_i > 0$.  
Numerical issues will arise if $b_i$ is non-positive, causing the exponent to diverge for large $q$. 
To constrain $b_i$, it is defined by a ``softplus'' transformation of unconstrained $b^*_i \in \mathbb{R}$: 

\begin{align}
    b_i &\equiv \frac{1}{\gamma_b}\log(1 + \mathrm{e}^{\gamma_b b^*_i}),
\end{align}

\noindent where $\gamma_b > 0$.
The value of $\gamma_b$ is chosen to be large enough such that for typical values, $b_i \approx b^*_i$.
The same constraint was found to be unnecessary for $a_i$.

To create the starting seeds, the median of each of the three timing parameters from a previous calibration were found. 
The same three parameter values were then used to identically initialize each PMT.

\subsection{Time Residuals}

PMT hit times, $t^i_\mathrm{hit}$, are strongly and systematically dependent on the event vertex position.
The hit times are also measured relative to the global trigger time, which occurs at an arbitrary offset from the actual event time, $t_\mathrm{event}$. 
Assuming the event is point-like compared to the detector resolution, the remaining random element, primarily relating to the light emission time profile and PMT time jitter, is thus characterised by the time residual for each PMT hit $i$:

\begin{equation}
    t^i_\mathrm{res} = t^i_\mathrm{hit} - t_\mathrm{event} - t^i_\mathrm{TOF},
\end{equation}

\noindent for PMT hit time $t^i_\mathrm{hit}$, event time $t_\mathrm{event}$ and time of flight from vertex $t^i_\mathrm{TOF}$.

For the correct position and event time, the time residual distribution across the hit PMTs should be sharply peaked, but with a long tail at late times due to scintillator reemission.
Any errors in the PMT timing calibration or vertex reconstruction will then broaden this distribution.
This forms the heart of the approach; individual events are assumed to produce light emanating from a single position that varies from event to event, while calibration parameters remain constant over the entire collection of events.
Individual event positions are reconstructed using candidate calibration parameters to subsequently calculate the time residuals.
The event positions and calibration constants can be adjusted until the time residual distributions are sufficiently narrow at their peak.
To do this, we make a fit to the time residual distributions with a simple model that has a sharp peak but with a long tail to account for the scintillator reemission.
Using this model as a log likelihood loss function, the event positions and calibration parameters can be optimized simultaneously using gradient descent methods.
Since any errors in the candidate calibration parameters may lead to biases in the position reconstruction, the simultaneous vertex and calibration optimization is crucial.

For this calibration, we use \ce{^210Po} alpha decays with $Q = \qty{5.4}{\mega\electronvolt}$ \cite{shamsuzzohabasuniaNuclearDataSheets2014} which gives approximately \qty{0.45}{\mega\electronvolt} of visible energy ({\em i.e.} the equivalent energy of an electron that would produce the same amount of light) due to nuclear quenching. 
This is an abundant background in the SNO+ detector due to radon contamination and can easily be identified by an energy cut.  
The $\alpha$ particles leave micron length energy depositions in the scintillator providing a point-like light source for calibration.

\begin{figure}[htb]
    \centering
    \includegraphics[width=0.5\columnwidth]{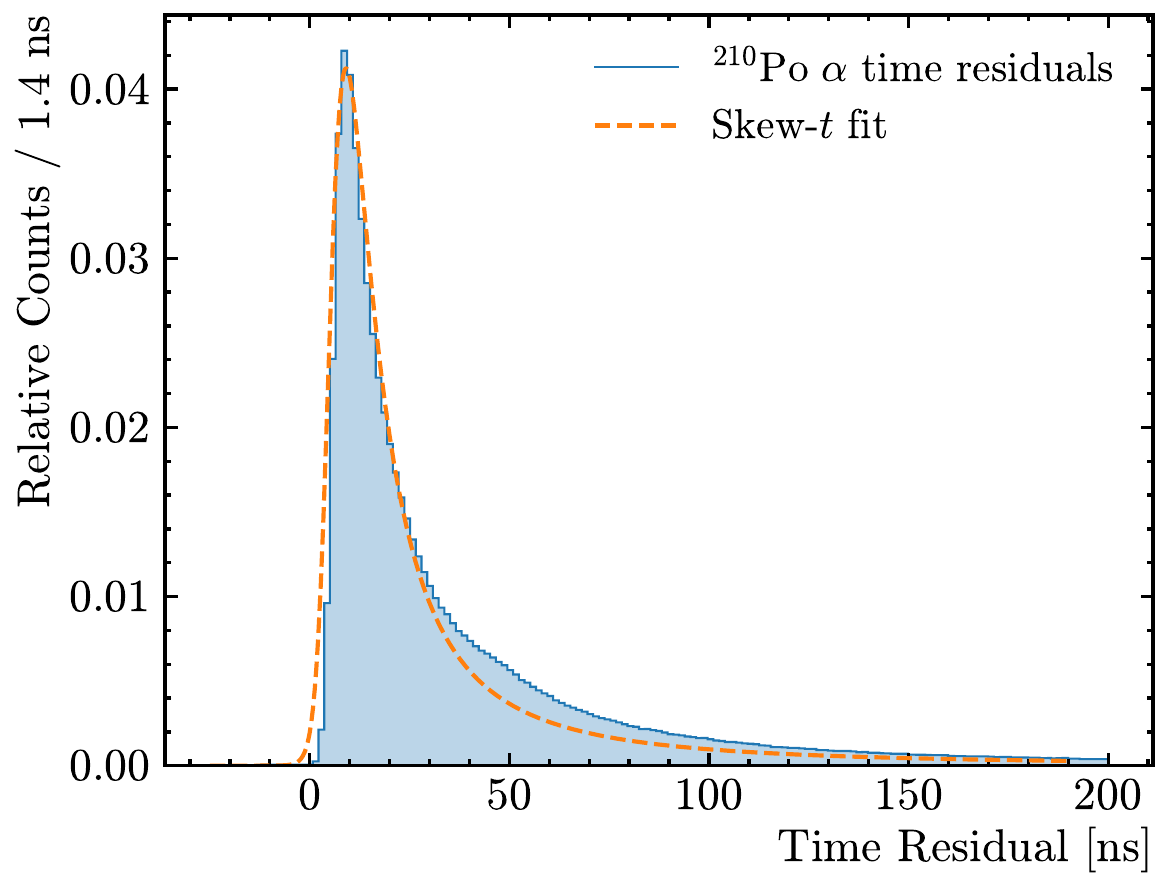}
    \caption{Time residual distribution for \ce{^210Po} decays from MC. The Jones and Faddy skew-$t$ distribution is fit to the distribution and used as a loss function during training.}
    \label{fig:time_residual_distribution}
\end{figure}

The time residual distribution for \ce{^210Po} decays from Monte Carlo (MC) simulations is shown in \autoref{fig:time_residual_distribution}.
This distribution needs to be modelled so that it can be used as a loss function for gradient based optimization.
The Jones and Faddy skew-$t$ distribution, $f(x)$, was chosen with scale and shift parameters $\mu$ and $\sigma$, and parameters that adjust the skew, $\alpha$ and $\beta$ \cite{jonesSkewExtensionDistribution2003}:
\begin{equation}
     f(x) = A \left(1+\frac{x}{\sqrt{\alpha+\beta+x^2}}\right)^{\alpha+1/2} 
     \left(1-\frac{x}{\sqrt{\alpha+\beta+x^2}}\right)^{\beta+1/2} \nonumber,
\end{equation}
with normalizing constant and beta function $B(\alpha,\beta)$
\begin{equation}
    A = \frac{1}{2^{\alpha+\beta-1}B(\alpha,\beta)(\alpha+\beta)^{1/2}},
\end{equation}

and normalized time residual

\begin{equation}
    x = \frac{t_\mathrm{res} - \mu}{\sigma}.
\end{equation}

This is an extension of the Student's $t$-distribution, giving heavy tails, and allowing for non zero skew.
With fit parameters fixed during training, the log likelihood is also simple enough to be used as a loss function.
A convenient feature of this distribution is that it has support on the entire real line, so no discontinuities arise during training if an anomalous time residual is calculated.

The fit is shown in \autoref{fig:time_residual_distribution} and describes the distribution well enough for our purpose.
In general, the precise nature of this fit will not significantly affect the end calibrations.
Each PMT channel will sample across the entire time residual profile during the calibration so any bias introduced to a PMT by a poor fit will be identical across all the PMTs. 
Since we can only measure relative PMT hit times, this does not make a difference to the calibration. 
Due to the complexity of the calibration procedure, propagating the error in the time residual fit to the final timing calibration is non-trivial.
However, we can roughly test the sensitivity by redoing the studies in \autoref{sec:monte_carlo} with $\sigma$ scaled by a factor of two in the fit.
This lead to a 30\% decrease in accuracy compared to \autoref{fig:mc_time_walk_residuals} which is still well within the requirements for a good calibration.

We use constant effective speeds of light in scintillator and water of $v_\mathrm{scint} = \qty{183.5}{\milli\metre\per\nano\second}$ and $v_\mathrm{water} = \qty{217.55}{\milli\metre\per\nano\second}$,
respectively. 
These velocities were taken from earlier detector calibrations. 
The value of $v_\mathrm{scint}$ was also tuned to reduce the PMT $z$ position dependence of the delay residuals in \autoref{fig:mc_delay_residuals}.
Our calibration is not sensitive to small variations in these values.
In principle, they could also be floated in the optimization, however we found these had weak convergence properties and did not seem to offer much advantage.

The time of flights are calculated assuming a simplified straight line path despite the high effective refractive index difference between scintillator and water (approximately 1.62 and 1.38, respectively).
Through MC tests, it was found that the error in the time of flight from the straight line and constant effective velocity assumptions is small. 
Comparing the median time of flight error with the angle that the light path makes to the scintillator-water interface normal, the deviation is less than \qty{0.1}{\nano\second} across all angles. 
Since the calibration averages over a wide variety of event positions inside the detector, this error is averaged out. 
If any effect remains, this should be seen as a notable systematic bias in the calibration, but as shown in \autoref{fig:mc_delay_residuals} no significant bias is observed in MC.
Future work could attempt more accurate time of flight calculations, but in this case we do not expect substantial improvements.  

To achieve good timing resolution, a large dataset is needed to sufficiently sample the time residual distribution using hits from all the PMTs.
This is provided by the abundant \ce{^{210}Po} background in SNO+.
In our case, datasets of around 7 million \ce{^{210}Po} decays were used, collected over the space of 6 days giving around 100,000 hits per PMT.
Due to the large number of hits per PMT, the simple per PMT timing model in \autoref{eq:time_walk_model} and the fact that each PMT has independent characteristics from each other, no regularization was used. 
The most frequent the PMT timing calibration was applied in SNO/SNO+ was monthly, so 6 days is a sufficiently short time period to ensure the calibration does not change significantly.

Trying to simultaneously optimize the vertex position and time of each event along with the PMT timing calibration constants is a high dimensional problem and would be extremely challenging to perform directly. 
Instead of floating the position and time of each event as parameters to be fit, a neural network is trained to reconstruct the event vertex from the PMT hit times. 
The neural network contains approximately 900,000 parameters which is much smaller than the 4 $\times$ 7 million parameters for the position and time needed for a direct fit and can easily scale to datasets of any size without increasing the number of parameters.
The uncalibrated hit times are corrected for by the current candidate calibration parameters with these parameters being updated during the course of the calibration.
These candidate calibrated hit times are then fed into the neural network to predict the position and time of each event.
The time residuals are calculated using these candidate calibrated hit times in conjunction with the predicted position and time through a straight line of flight calculation.

The log likelihood given by the skew-$t$ distribution (with parameters previously fit from MC) is then used to construct the loss function.
The neural network and the timing parameters can then be optimized simultaneously through minimization of this loss function to yield the best timing model parameters. 
In this way, the reconstruction neural network does not need to be trained on explicit MC truth labels, as is typical, and the calibration directly trains it on data in a form of unsupervised learning.
This allows for a simple calibration procedure that can be performed end-to-end and requires minimal dependence on MC simulations.





\subsection{Model Architecture}

The core of the deep learning model is a transformer encoder \cite{NIPS2017_3f5ee243} based on earlier work on position reconstruction in SNO+ \cite{hewittMachineLearningEvent2025}. 
The same model hyperparameters were used from the original development on position reconstruction.
Transformers are powerful neural networks that use self-attention to learn relationships between different inputs.
They are a natural choice for a scintillation detector with events that produce a variable number of PMT hits.
Transformers have an $O(n_\mathrm{hits}^2)$ complexity. 
However, each event has less than 180 hits and with the easily accessible, highly optimized implementation of transformers this is a computationally efficient architecture to train in the time scale we need.


Each PMT hit forms an individual token, so the self-attention mechanism acts between every pair of hits.
The input features to each token are the PMT ID and the hit time.  
Due to the presence of an arbitrary global time offset in the hit times, only relative hit times should be used by the model.
To enforce this, the hit times are centered by the mean hit time of the event, ensuring that the model is invariant to any global shift.
The original transformer was applied to token sequences in which each token had its ordinal number encoded.
It is not obvious how meaningful sequences could be constructed from PMT hits. 
The hits could be ordered by hit time, but PMT hit times are not evenly spaced, so this would not fully capture the timing information. 
Therefore, no sequence position encoding is used to maintain hit permutation invariance.

To construct each token, each PMT is given an ID (from \autoref{eq:pmt_id}) and each ID is given an embedding of dimension 64 that is learned during training. 
During previous development for reconstruction, this was found to give better results compared to embedding the position of the PMT.
The hit time is passed through a multilayer perceptron with a $\tanh$ activation and a single 512 dimension hidden layer to give a matching length 64 embedding.
The final token is then formed by summing these two embeddings together.

The transformer block has 6 layers, each with 4 attention heads and a feed-forward hidden dimension of 128. To extract a global representation of the event, the output of the transformer is passed through a mean pooling layer to give a single vector of length 64 for the entire event. 
A linear layer is then applied to give the 4 outputs of the model: the $x$, $y$ and $z$ positions and the time $t$ of the event.

\subsection{Calibration Procedure}

The calibration procedure is as follows:
\begin{enumerate}
    \item Batches of \ce{^210Po} events are retrieved with the PMT hit charges $q_i$ and uncalibrated times $t^i_\mathrm{uncal}$.
    \item These times are corrected according to the candidate PMT timing models: $t^i_\mathrm{cal} \equiv t^i_\mathrm{uncal} - \tau_i(q_i)$.
    \item The $t^i_\mathrm{cal}$ and PMT IDs are passed to the transformer model to reconstruct the event position and time.
    \item Using a straight line time of flight calculation the time residuals, $t^i_\mathrm{res} = t^i_\mathrm{cal} - t_\mathrm{event} - t^i_\mathrm{TOF}$, are calculated.
    \item The loss is calculated as the negative log likelihood of the time residuals given the Jones and Faddy skew-$t$ distribution fit to the MC time residual distribution in \autoref{fig:time_residual_distribution}.
    \item Using the loss, gradients are calculated on the PMT timing model and the transformer model parameters.
    Since $t^i_\mathrm{cal}$ is used as input into the transformer model and directly used in the time residual calculation, gradients are allowed to flow to the timing model in two different paths: one passing through the neural network and one directly.
\end{enumerate}

The calibration is performed using a batch size of 2048 events, using the Adam optimizer \cite{kingmaAdamMethodStochastic2017} and a fixed 500,000 batches.
Using 100,000 batches or decreasing the batch size to 1024 only lead to a 10\% decrease in timing accuracy in the MC tests from \autoref{sec:monte_carlo}.
The 1cycle learning rate scheduler \cite{smithSuperConvergenceVeryFast2018} is used to vary the learning rate from an initial value of \num{2e-5} to a maximum of \num{1e-3} at the first 30\% of steps and then decreases to \num{5e-8} at the last step.
The maximum learning rate was coarsely chosen as a balance between training time and performance.
The end timing calibration are not very sensitive to these choices.
Doubling the maximum learning rate lead to no appreciable difference in timing accuracy. 

The model used for the timing in \autoref{eq:time_walk_model} is a simple exponential model, so it benefits greatly from the small learning rate at the end of training given by the 1cycle scheduler to fine tune the parameters.
To this end, the calibration constants should always be taken from the final step of the calibration.
If the transformer model overfits at the end of the calibration then that may reduce the calibration accuracy.
The calibration is not very sensitive to a small increase in the position or time reconstruction resolution since the calibration averages over a large number of events.
However, any bias to the position or time reconstruction would lead to a corresponding bias in the calibration constants.
For this reason, it is important to ensure that the the model does not overfit by monitoring a validation loss. 
Across the many different training runs performed, no significant overfitting was observed in the validation loss, so this does not seem to be a problem in practice.

The model is implemented in PyTorch with the FlashAttention-2 \cite{dao2023flashattention2fasterattentionbetter} implementation of self attention.
On a single NVIDIA H100 \qty{96}{\giga\byte} GPU the calibration takes approximately 6 hours to complete.

\section{Validation}
\subsection{Event Selection}\label{sec:event_selection}
Events near the AV will have a larger number of paths that have large incident angles at the scintillator-water interface.
This will mean the angle of refraction will be large and even may undergo total internal reflection which are both not modelled well by the straight line time of flight calculations.
Therefore, a fiducial cut is made to remove events near the AV with a reconstructed radius (with respect to the centre of the PMTs) greater than \qty{4}{\metre}. 
Additionally, \ce{^{210}Po} rates are higher near the top of the detector on the AV edge, so the cut makes the event distribution more uniform. 
Since this cut relies on a position reconstruction that is itself reliant on a previous PMT timing calibration, this might raise a concern of a possible bias being introduced.
To test the sensitivity of the calibration to this, MC studies were performed in which an imposed initial $z$ bias in the fiducial volume selection as large as \qty{20}{\centi\meter} were introduced. 
This lead to a $z$ timing bias of \qty{80}{\pico\second} in magnitude comparing the top and bottom of the detector.
The largest reconstruction bias from this timing bias would be achieved if only hits from the top and bottom of the detector were used in reconstruction leading to a $z$ bias of $\qty{80}{\pico\second} \times v_\mathrm{eff} / 2 \sim \qty{8}{\milli\metre}$ with effective speed of light $v_\mathrm{eff} \sim \qty{200}{\milli\meter\per\nano\second}$.
In standard laserball calibrations we expect a bias of around \qty{10}{\centi\metre} in magnitude.
If desired, even this could be eliminated with iteration.


The data and simulations used in this analysis use runs immediately after completion of the SNO+ liquid scintillator fill in May 2022.
This run range was chosen due to the high rate of \ce{^{214}BiPo} coincidences from the elevated radon contamination from the fill operation.
These can provide a useful dataset for comparing timing calibrations as performed in \autoref{sec:bipo}.
MC studies were based on generated \ce{^{210}Po} decays in the detector.
For data studies, events were selected with reconstructed energies between \qty{0.35}{\mega\electronvolt} and \qty{0.55}{\mega\electronvolt} to select \ce{^{210}Po} decays.
This cut gives an approximately \qty{98}{\percent} pure \ce{^210Po} sample.
The other significant backgrounds we expect in this region are the beta emitters \ce{^210Bi} and \ce{^39Ar}.
Since the salient features of \ce{^{210}Po} alpha decays are point-like energy depositions, contamination with beta events that are also very point like should not significantly affect the calibration.
Both data and MC datasets use approximately 7 million events after cuts.

\subsection{Monte Carlo Validation}\label{sec:monte_carlo}
Within the assumption of the timing model in \autoref{eq:time_walk_model}, MC can explicitly validate the timing accuracy of the entire calibration method and check for any systematic effects caused by insufficient physical modelling. 
The calibration can be applied to MC that has been ``uncalibrated'' by known timing models allowing a direct comparison between calibrated and truth timing models.
The timing models used are the same as those found after calibrating on data in \autoref{sec:data_validation}.
Uncalibration for each PMT is performed by taking the truth timing model for the PMT, finding the timing model value at the hit charge, and then subtracting the MC truth hit time with this timing model value. 
The calibration process is performed on this uncalibrated MC sample and the fitted timing models can be compared to the ground truth timing models that were used to uncalibrate. 

\begin{figure}[htb]
    \centering
    \includegraphics[width=0.5\columnwidth]{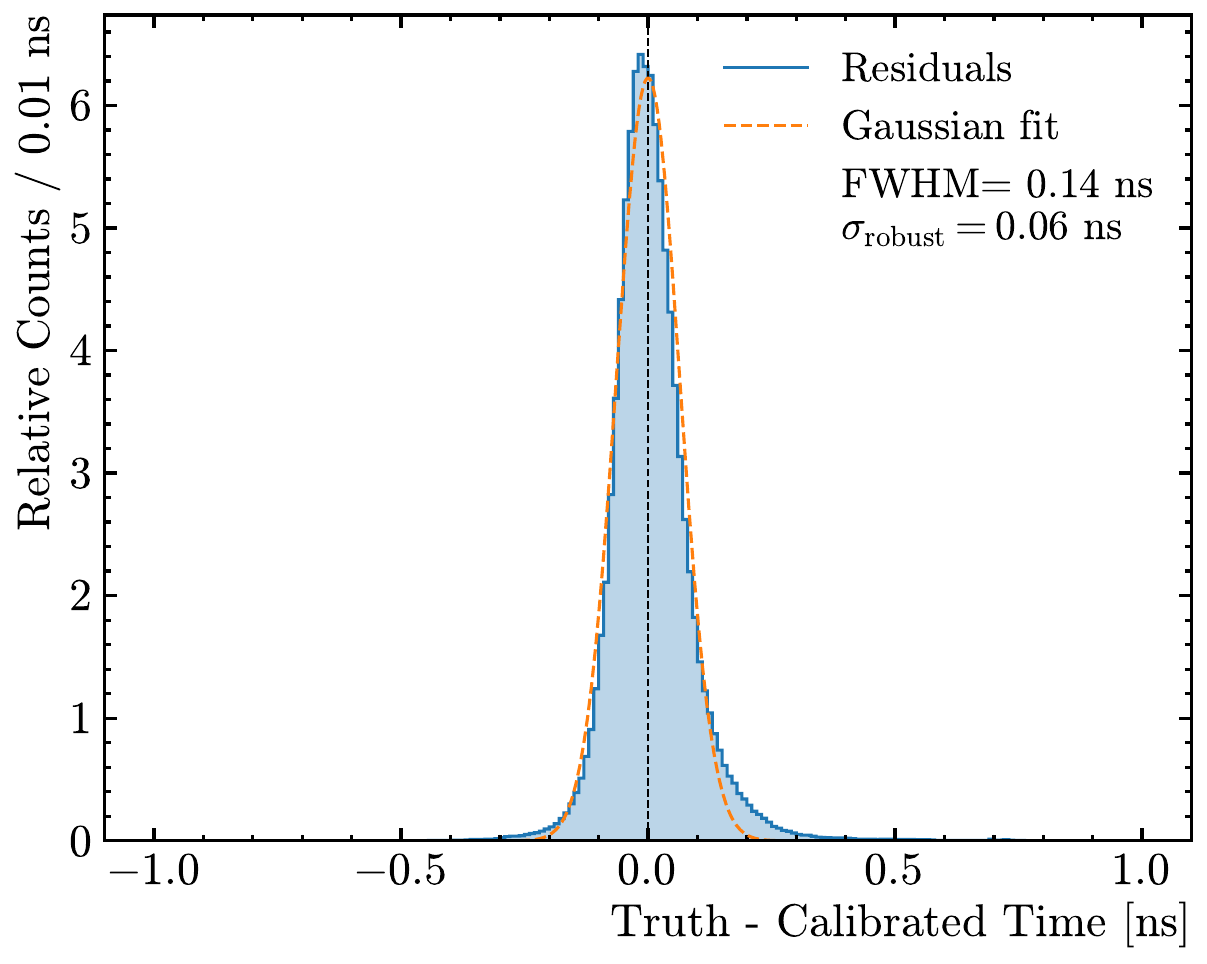}
    \caption{Distributions of PMT timing model residuals over all PMTs. The standard deviation of the residuals is estimated by taking the interquartile range and dividing by \num{1.35} for a robust estimate.}
    \label{fig:mc_time_walk_residuals}
\end{figure}

The residual of interest to evaluate the accuracy of the timing model for PMT $i$ is given by 

\begin{equation}\label{eq:weighted_time_walk_residual}
    \tau_i^\mathrm{fit}(q_i) - \tau_i^\mathrm{truth}(q_i),
\end{equation}

\noindent for charge $q_i$ where $\tau_i^\mathrm{truth}$ is the MC truth timing model and $\tau_i^\mathrm{fit}$ is the fitted timing model from the calibration method.
Each PMT will have a different distribution of charges. Thus, in order to get a distribution of \autoref{eq:weighted_time_walk_residual} over all PMTs, \num[group-separator={,}]{10000} samples of $q_i$ for each PMT are drawn from the corresponding PMT's charge histogram, from which the above residual is calculated and binned. Since reconstruction is invariant to a global PMT time offset, the distribution is centred by the median of the distribution.
The results are shown in \autoref{fig:mc_time_walk_residuals}.

\begin{figure}[htb]
    \centering
    \includegraphics[width=0.6\columnwidth]{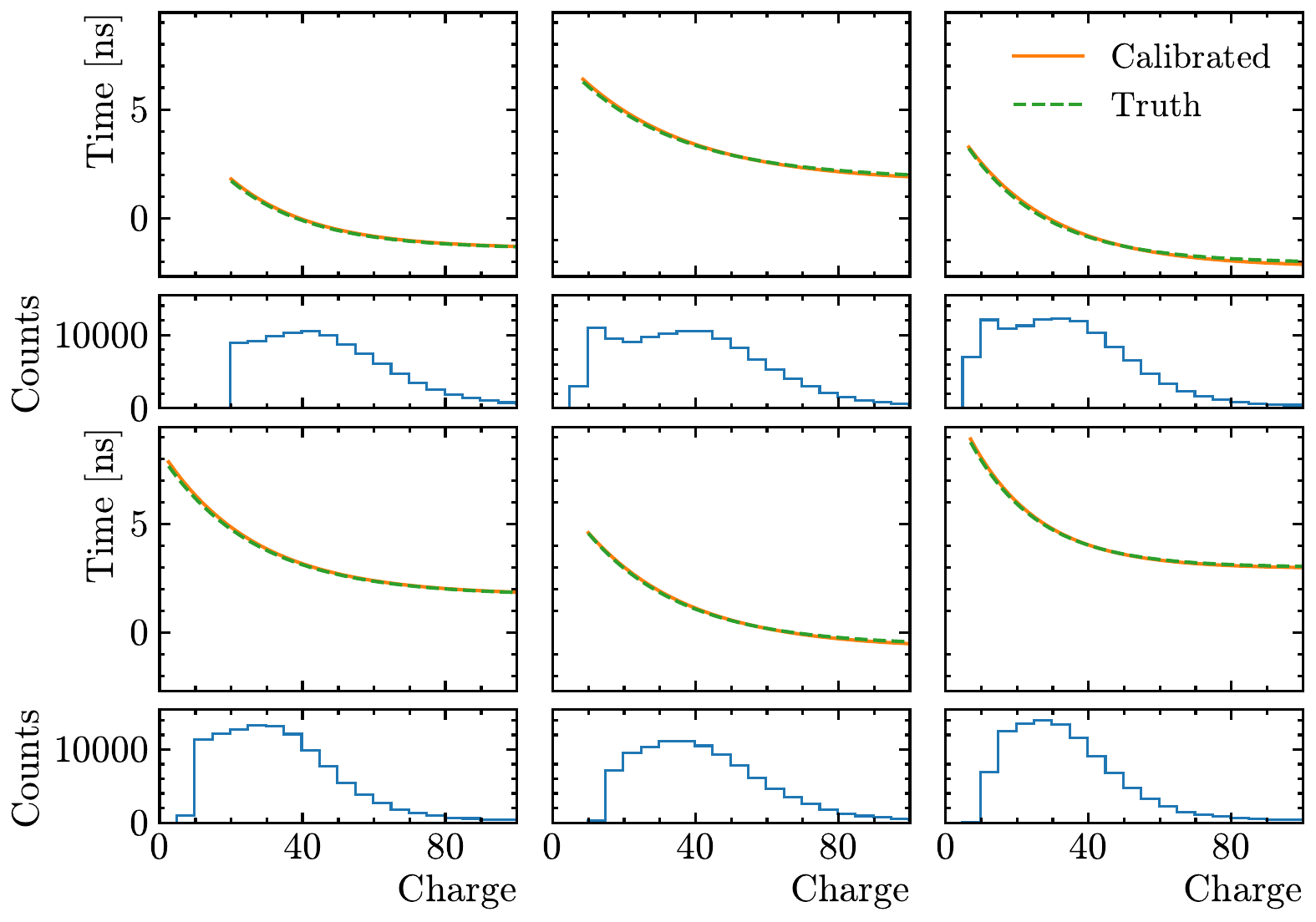}
    \caption{Examples of PMT timing models from MC truth and as fitted by the calibration method. The inset histogram shows the distribution of charges for this PMT. 
    Time walks are only shown for charge bins which have non-zero counts.
    The height of the flat part of the curve gives the cable/electronic delay.
    }
    \label{fig:mc_time_walk_examples}
\end{figure}

With a full width half maximum (FWHM) of \qty{0.14}{\nano\second}, this is much smaller than the transit time spread of \qty{1.5}{\nano\second} (RMS) for the Hamamatsu R1408 PMTs used in SNO+.
The distribution of the residuals is skewed positively, which is likely due to the difficulty of fitting the timing model at the steep, low charge part of the spectrum, as shown in \autoref{fig:mc_time_walk_examples}.

\begin{figure}[htb]
    \centering
    \begin{subfigure}{0.48\columnwidth}
        \centering
        \includegraphics[width=1.0\columnwidth]{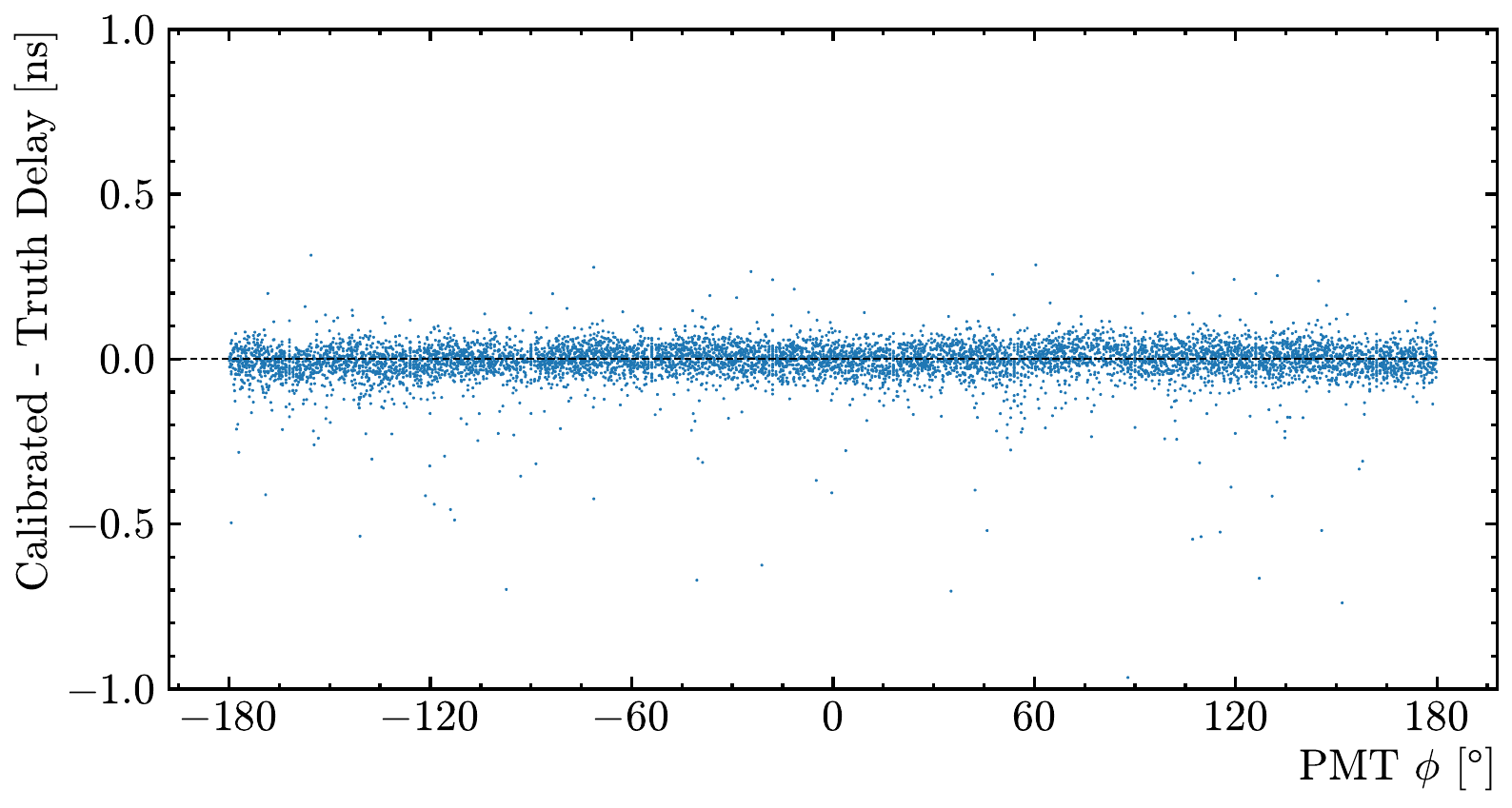}
    \end{subfigure}%
    \hfill
    \begin{subfigure}{0.48\columnwidth}
        \centering
        \includegraphics[width=1.0\columnwidth]{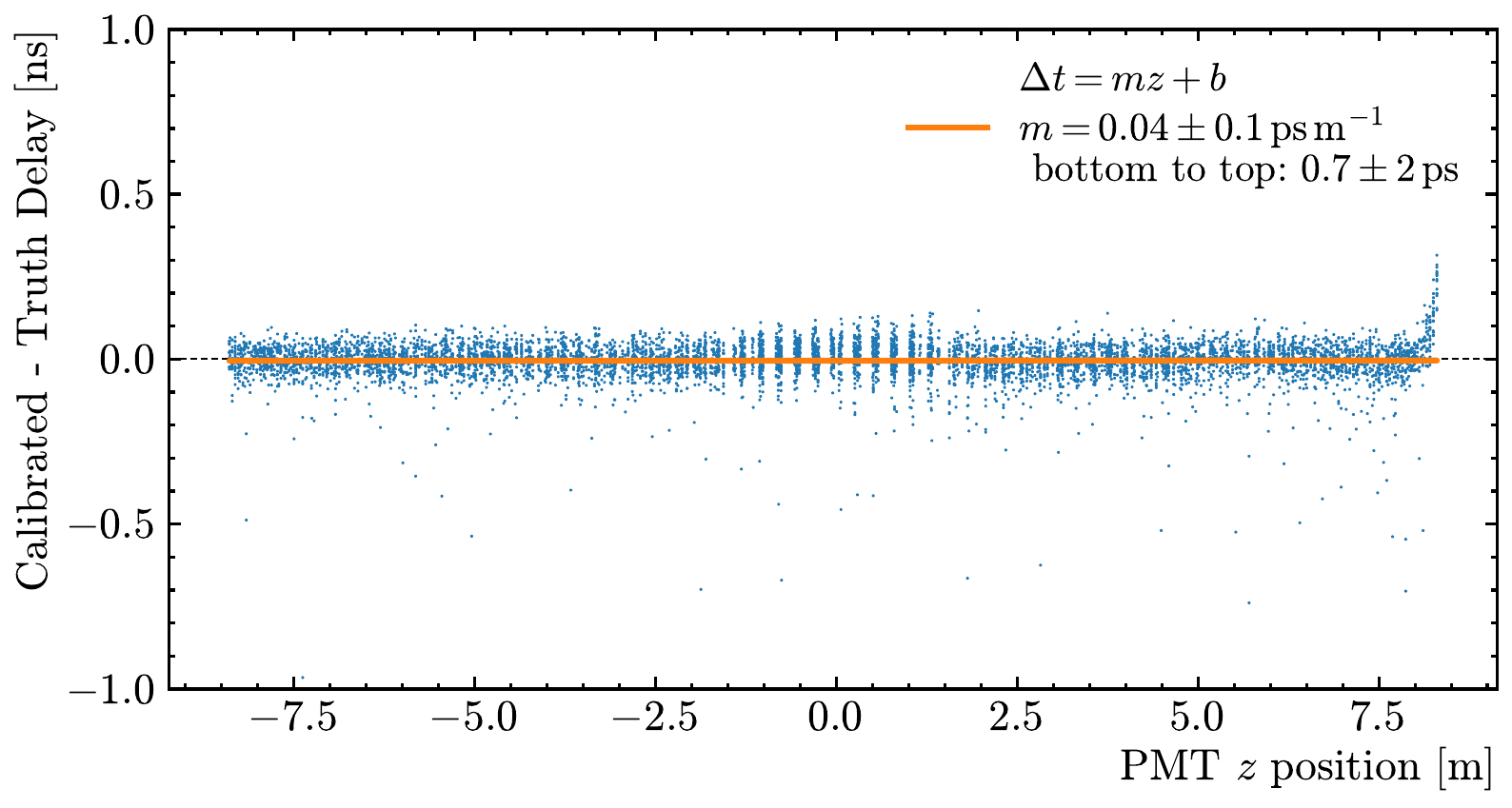}
    \end{subfigure}
    \caption{Each blue dot represents the delay residual for a PMT. 
    For the PMT $z$ position plot, a linear fit is shown in orange to illustrate the small dependence on $z$ position. 
    The ``bottom to top'' shows the difference in delay residual from the bottom to the top of the detector based on the linear fit.}
    \label{fig:mc_delay_residuals}
\end{figure}

An important consideration is whether the timing calibration has any dependence on the spatial position of the PMTs that may be caused by insufficient physical models including the simple time residual fit and straight line time of flight calculations.
The delay residuals for each PMT are calculated by grouping the timing model residuals in \autoref{eq:weighted_time_walk_residual} by PMT and taking the median to average over the charge distribution.
The delay residuals are then plotted against the PMT azimuthal angle $\phi$ and $z$ position in \autoref{fig:mc_delay_residuals}.
These show no significant dependence on PMT $\phi$ or $z$. 
The large uptick at large $z$ is due to the neck of the AV increasing the light path length. 
This optical effect of the neck is not easily modelled (and is not modelled in the standard laserball calibration) so this feature is expected.
Between \qty{-1.5}{\metre} and \qty{1.5}{\metre}, there is also a small increase in delay (where there are large $z$ gaps between the PMTs), which is due to acrylic support plates attached to the AV that increase the path length. 

The calibration results also remain reproducible across different random seeds, new MC simulations and small changes to optical parameters.  

\begin{figure}[htb]
    \centering
    \includegraphics[width=0.5\columnwidth]{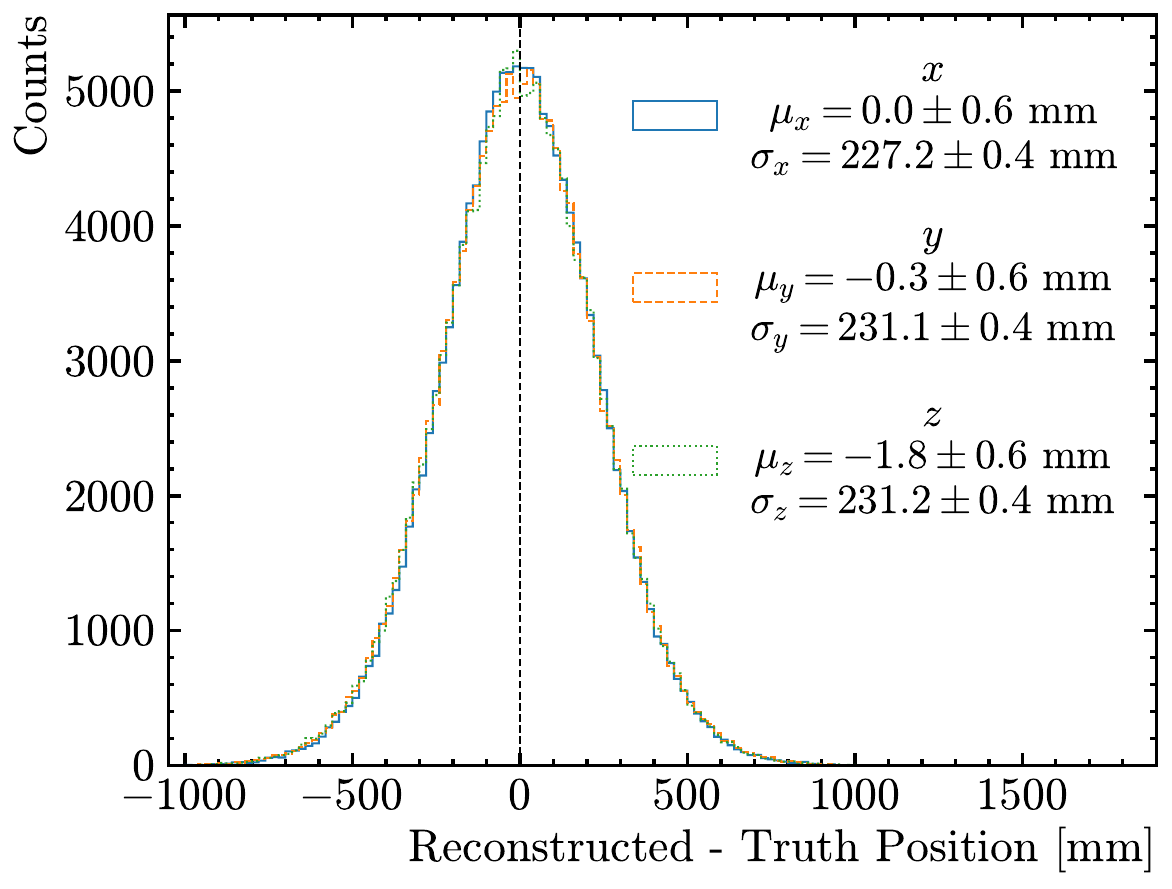}
    \caption{The position residuals (reconstructed - truth) as predicted by the position reconstruction neural network. The position reconstruction is carried out using PMT hit times determined by the calibration method described here.}
    \label{fig:pos_res_during_cal}
\end{figure}

Using a MC dataset not seen during calibration, the position reconstruction neural network was fed hit times calibrated by the procedure described here.
This tests the overall performance of the calibration and position reconstruction together.
Position residuals compared to MC truth positions are shown in \autoref{fig:pos_res_during_cal}.
Position reconstruction performs well during the calibration, with minor biases.
Given the simplified time residual and optical model used in the calibration, the resolution is surprisingly close to that of the standard likelihood based position reconstruction algorithms\footnote{In SNO+, standard position reconstruction uses a maximum likelihood method from probability distribution functions of time residuals derived from MC.} for such alpha particles ($\sim\qty{215}{\milli\metre}$). 
While efforts could be made to improve this resolution, we do not expect it to significantly improve the calibration. 
The calibration effectively averages over all the events in the dataset so the effect of the position reconstruction error will be extremely suppressed.


\subsection{Data Validation}\label{sec:data_validation}
Despite lacking ground truth timing models to compare to, some validation can still be made on real data.
These include: 1) looking at the electronic channel structure of the delays found, 2) comparing to the standard SNO+ laserball calibration, and 3) looking at the position resolution using \ce{^{214}BiPo} coincidences.

\subsubsection{Delays}

The front end electronics for the PMTs are grouped together into 19 different crates.
Each crate is subdivided into 32 cards that each have 16 PMT channels.
A unique PMT ID is given to each PMT based on its crate, card and channel number:
\begin{equation}\label{eq:pmt_id}
    \mathrm{ID} = 512 \times \mathrm{crate} + 32 \times \mathrm{card} + \text{channel}.
\end{equation}

Each PMT channel has a different electronic trace to traverse inside the crate and the cables should be the same length.
Thus, the delays for PMTs across crates should generally be independent, but, within a crate, there should be a consistent pattern with PMT ID.
This provides a characteristic structure that can be easily identified as a consistency check that the PMT calibration is performing as expected. 
The delays used here are given by the $c_i$ in \autoref{eq:time_walk_model}.

\begin{figure}[hptb]
    \centering
    \begin{subfigure}{0.5\columnwidth}
        \centering
        \includegraphics[width=1.0\columnwidth]{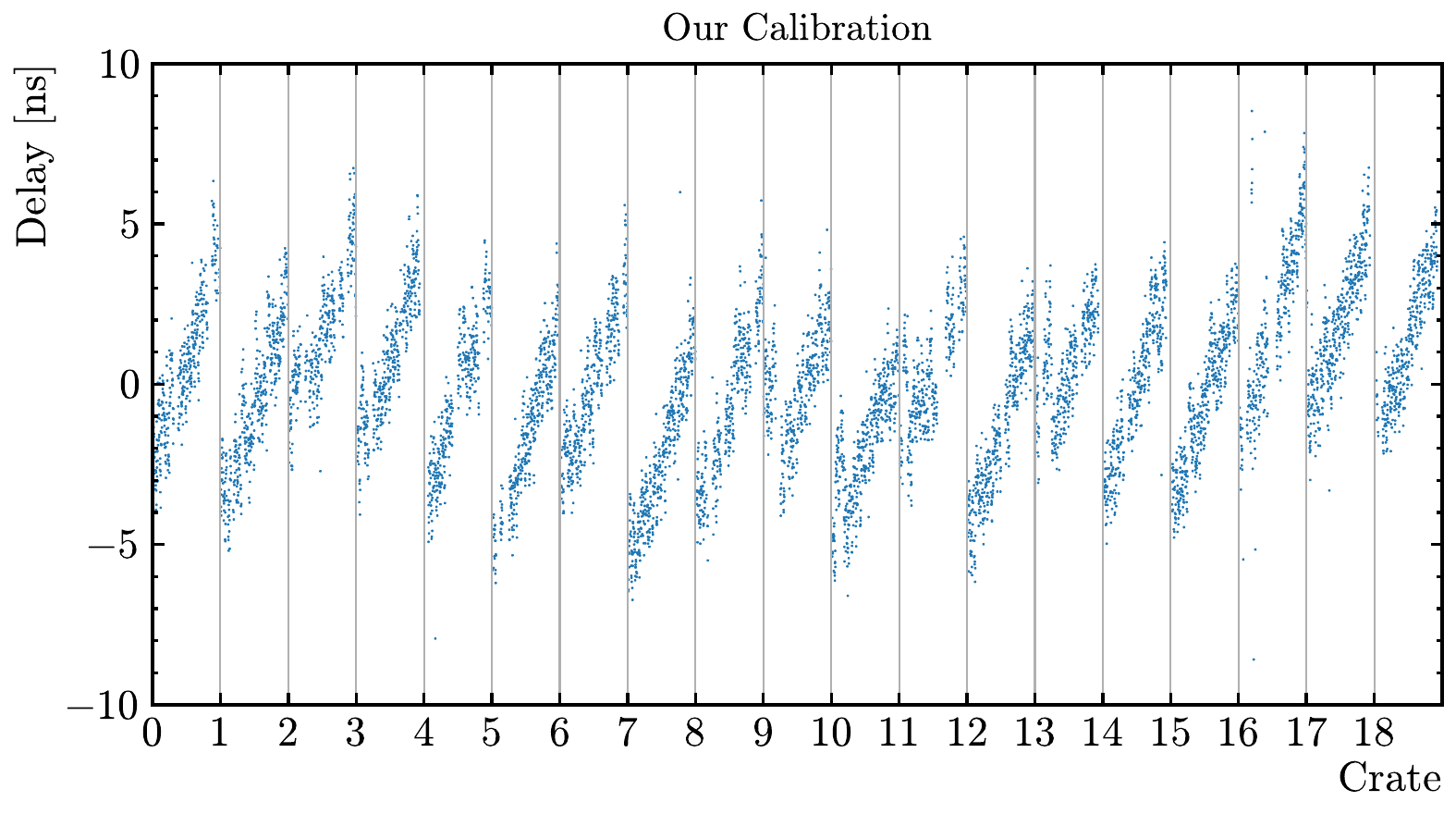}
        \caption{Our calibration delays}
        \label{fig:our_delays}
    \end{subfigure}%
    \hfill
    \begin{subfigure}{0.5\columnwidth}
        \centering
        \includegraphics[width=1.0\columnwidth]{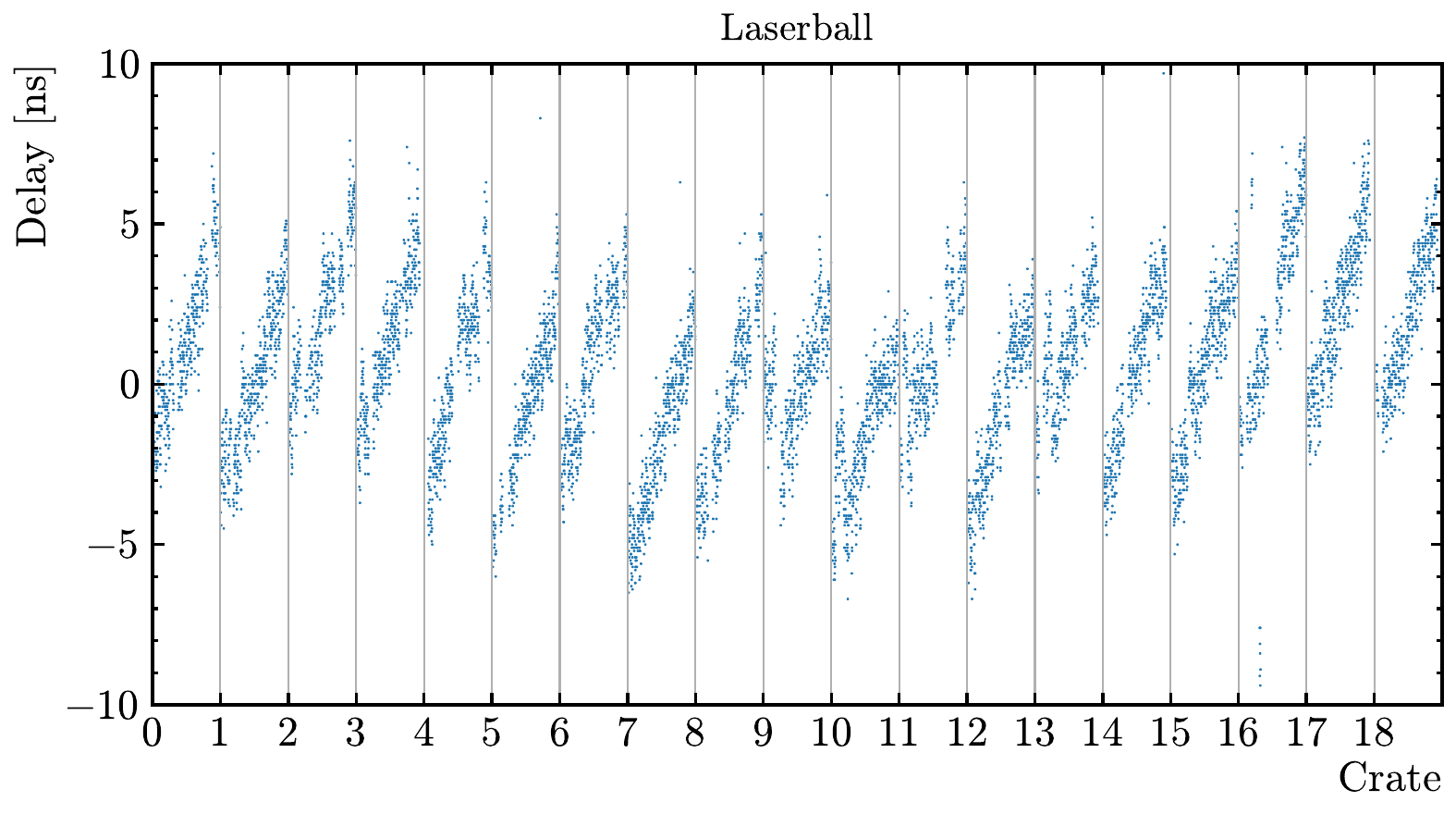}
        \caption{Standard laserball SNO+ calibration delays }
        \label{fig:standard_delays}
    \end{subfigure}
    \vfill
    \begin{subfigure}{0.5\columnwidth}
        \includegraphics[width=1.020\columnwidth]{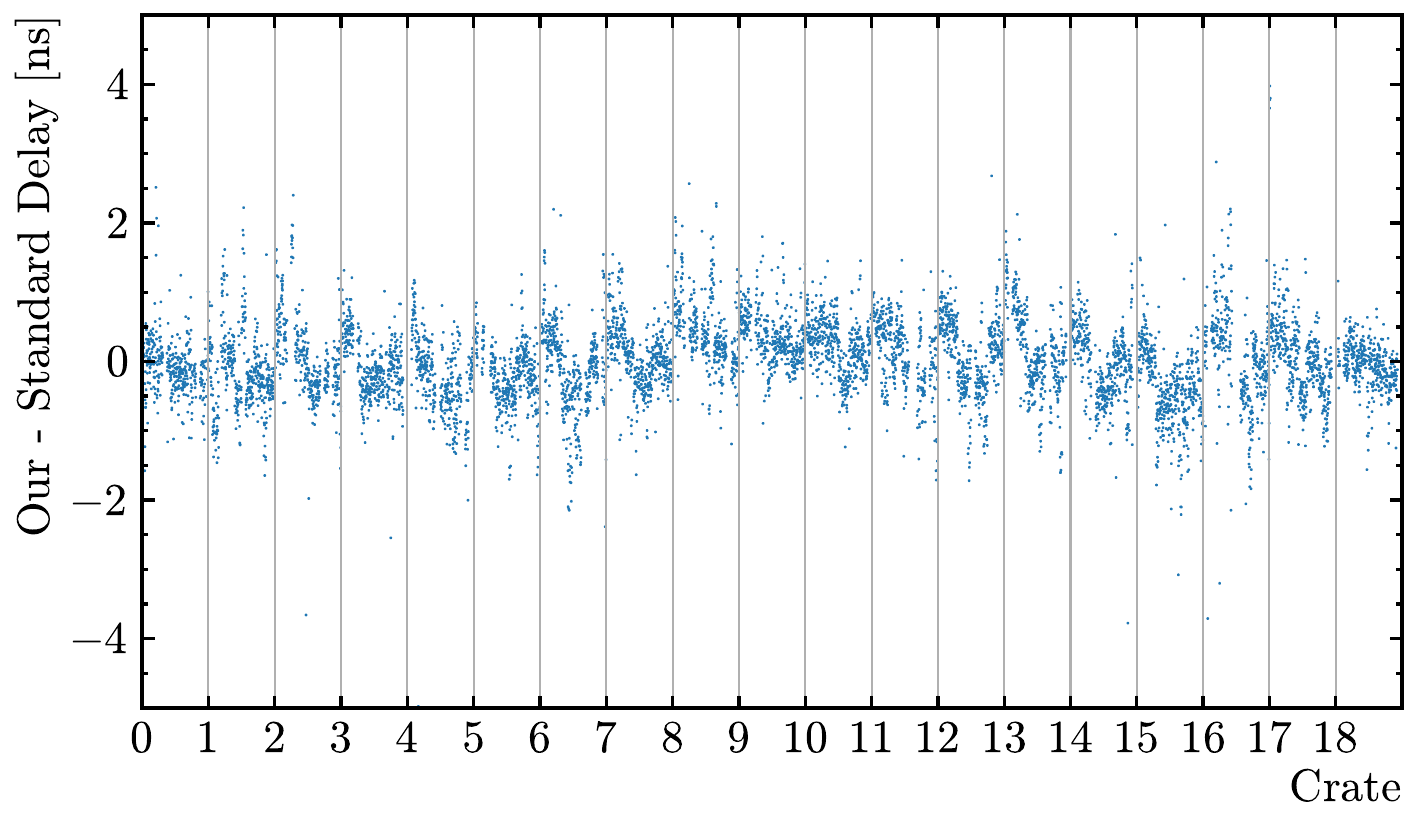}
        \caption{Residual between our calibrations and standard SNO+ calibration delays}
    \end{subfigure}
    \caption{Comparison of delays found by our calibration method and the standard SNO+ calibration using the laserball.}
    \label{fig:delay_comparison}
\end{figure}

\begin{figure}[hptb]
    \centering
    \includegraphics[width=0.75\columnwidth]{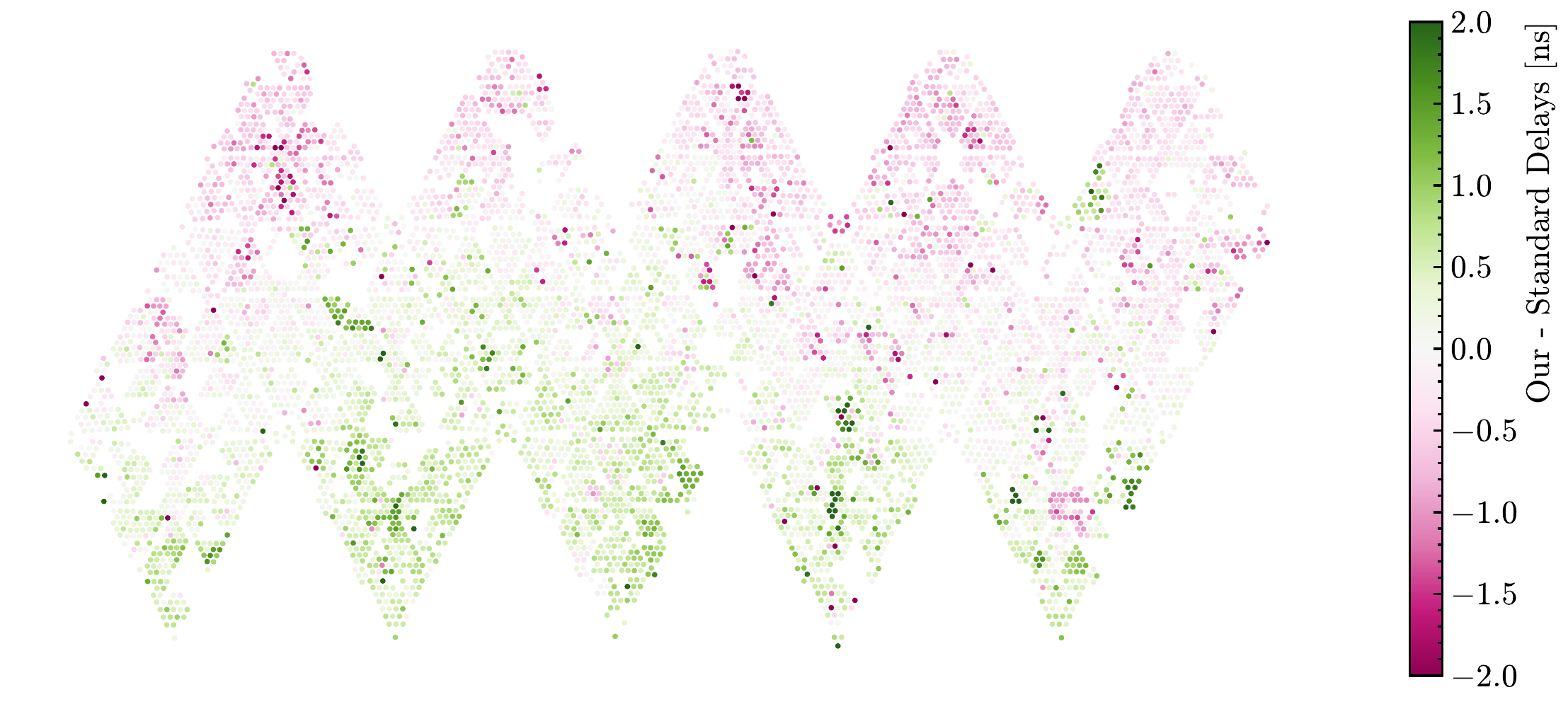}
    \caption{Residual between our calibrations and standard SNO+ calibration delays projected onto a flat map of the detector. The vertical $z$ direction runs from top to bottom and azimuthal angle $\phi$ runs horizontally. }
    \label{fig:delay_residual_flatmap}
\end{figure}

The structure of our delays in \autoref{fig:our_delays} show very similar crate by crate structure to the standard SNO+ delays \autoref{fig:standard_delays} \cite{descampsPCACalibrationSNO2016}.
The model has no knowledge of the crate/card/channel structure of the electronics (similarly to the laserball calibration) with the PMT IDs used purely as indices into lookup tables.
Thus, the replication of this structure indicates that the calibration is finding physically meaningful results. 

Comparing the delays from the two methods directly in \autoref{fig:delay_residual_flatmap}, there is a clear dependence of the delay residuals on the $z$ position of the PMTs.
This difference is approximately linear with a \qty{1}{\nano\second} difference from top to bottom of the detector based off the linear fit. However, this is consistent with the uncertainty in the absolute position of the laserball ($\sim \qty{10}{\centi\metre}$ \footnote{This calculation assumes a typical \qty{200}{\milli\metre\per\nano\second} speed of light leading to a $2 \times \qty{10}{\centi\metre} / \qty{200}{\milli\metre\per\nano\second} = \qty{1}{\nano\second}$ top to bottom timing difference.} 
) that was used for the initial timing calibration \cite{cameronPhotomultiplierTubeCalibration}. 

\subsubsection{\ce{^{214}BiPo} Coincidences}\label{sec:bipo}
In addition to the \ce{^{210}Po} decays used in our calibration, the intrinsic background from \ce{^{222}Rn} in the SNO+ detector provides a useful source of \ce{^{214}Bi}.
The \ce{^{214}Bi} decays to \ce{^{214}Po} via a $Q = \qty{3.27}{\mega\electronvolt}$ beta decay, shortly followed by an alpha decay with a \qty{164.3}{\micro\second} half-life in a sequence known as a \ce{^{214}BiPo} coincidence \cite{zhuNuclearDataSheets2021}.
The high $Q$-value of the \ce{^{214}Bi} decay and short half-life of the \ce{^{214}Po} decay allows these coincidences to be easily tagged.
The position resolution of the detector is $O(\qty{10}{\centi\metre})$, which is much larger than the typical distance the \ce{^{214}Po} nucleus will travel before decaying due to the short half-life.
However, it is still much longer than the \qty{400}{\nano\second} trigger window, so the individual \ce{^214Bi} and \ce{^{214}Po} decays will trigger separate events most of the time.

A purely data-driven metric of position resolution can thus be found by individually reconstructing the positions of the $\ce{^{214}Bi}$ and $\ce{^{214}Po}$ decays and finding the distance between the two reconstructed positions, denoted as $\Delta R \equiv \norm{\vec{x}_\mathrm{Bi} - \vec{x}_\mathrm{Po}}$.
Better timing calibrations results in better position reconstruction and, therefore, in smaller values of $\Delta R$.

\begin{figure}[htb]
    \centering
    \includegraphics[width=0.6\columnwidth]{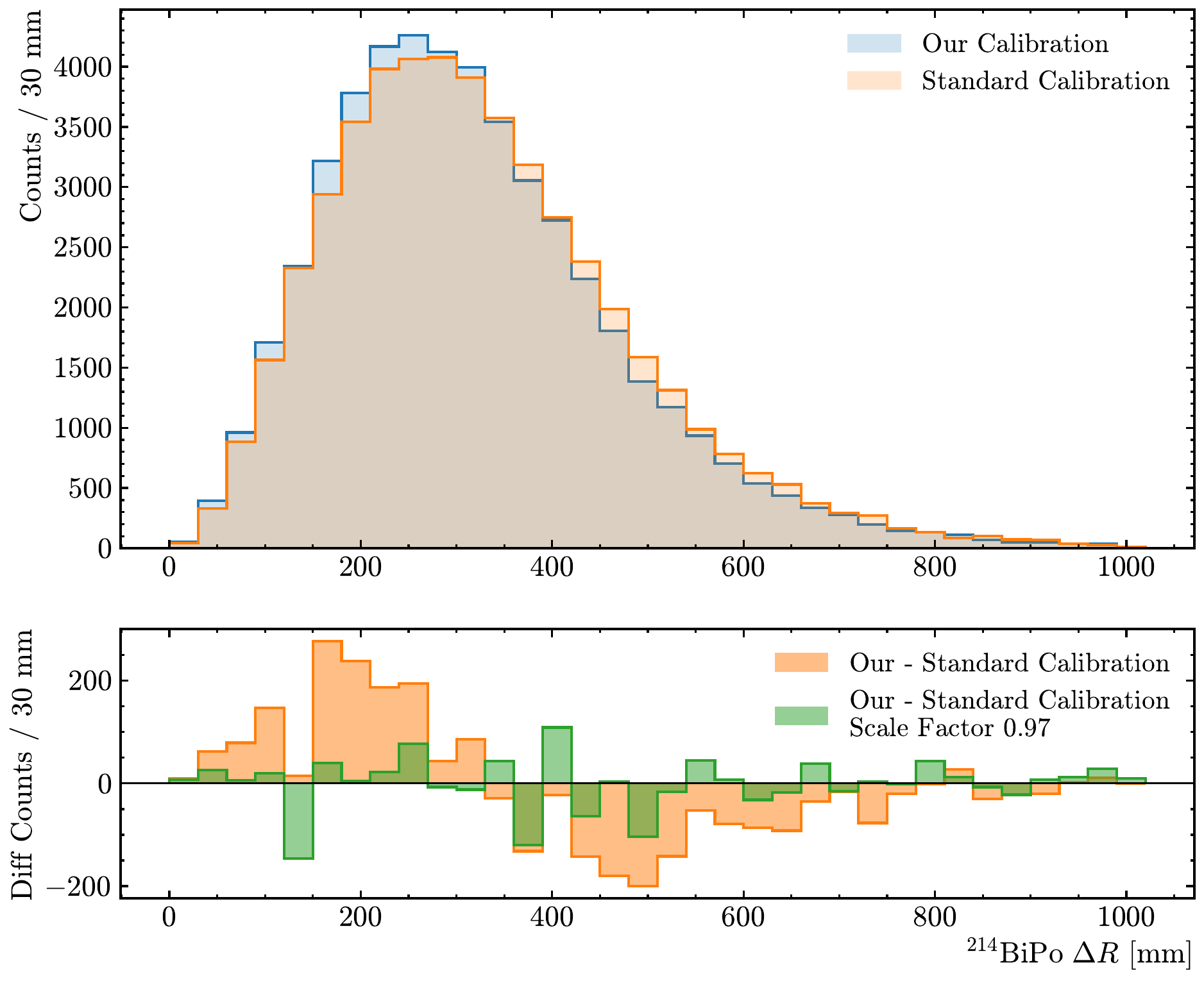}
    \caption{Distributions of $\Delta R$ for $\ce{^{214}BiPo}$ coincidences in data. Distributions are shown for standard SNO+ calibration and our calibration. Combined Bi and Po position resolution values are shown.}
    \label{fig:deltaR}
\end{figure}

The distribution resulting from the data-driven method of this paper is compared with that from the standard SNO+ laserball calibration in \autoref{fig:deltaR}. 
For this study, the same standard SNO+ likelihood-based position reconstruction is used for both, with results differing only as a consequence of which calibration time constants are used.
To compare the $\Delta R$ distributions from the two calibrations in a non parametric manner, the $\Delta R$ distribution from our calibration is measured as a scaled version, by factor $\alpha$, of the laserball distribution.
To find $\alpha$, we scaled the $\Delta R$ values from the laserball calibration to minimize the Kolmogorov–Smirnov test statistic between the two $\Delta R$ samples. 
This gives $\alpha = \num{0.97}$ showing that this calibration provides a slight improvement to the laserball calibration.
The smaller $\Delta R$ values can also be clearly seen in the difference plot in \autoref{fig:deltaR} where our calibration has more mass at lower $\Delta R$ and less mass at higher $\Delta R$ compared to the standard laserball calibration.



\subsubsection{Detector Monitoring}

\begin{figure}[htb]
    \centering
    \includegraphics[width=0.6\columnwidth]{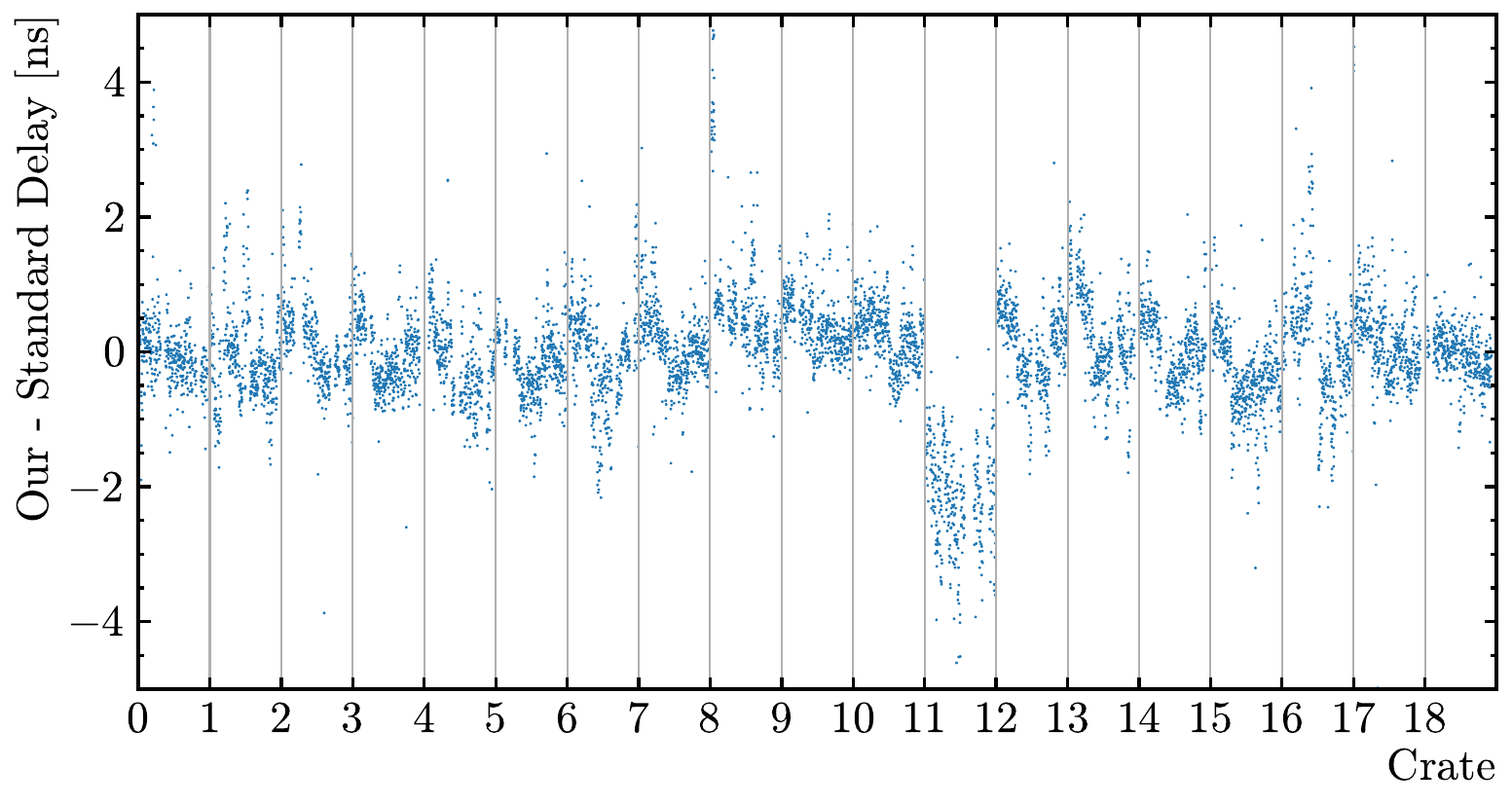}
    \caption{Delay residuals between our calibration and the standard SNO+ calibration for dataset taken before a lower level electronics calibration was performed.}
    \label{fig:bad_eca}
\end{figure}

Due to the ease of performing this calibration, it can be used to frequently monitor the stability of PMT timing calibrations.
The first attempt at performing this calibration on the same dataset in \autoref{fig:delay_comparison} showed significant deviations in the delays (defined by the $c_i$ in \autoref{eq:time_walk_model}) for crate 11, as shown in \autoref{fig:bad_eca}. 

This behaviour was entirely unexpected, but our calibration method is blind to crate segmentation and, once we were alerted to this, the effect could clearly be seen in the raw time residuals. Following a lower level electronics calibration\footnote{This lower level calibration involves finding the zero of the PMT charge values in addition to fitting the ADC counts of the capacitors to physical times. The calibration is performed with in situ electronics.}, performed shortly after this dataset was taken, a repeated application of our calibration indicated that this deviation had disappeared.
While the exact cause of this is still under investigation, the shift was clearly real and our calibration allowed us to identify when the issue occurred between normal calibration runs.
This also highlights the utility of this approach as a continuous monitor of detector performance and an aid to diagnosing and correcting potential issues.

Except for \autoref{fig:bad_eca}, all other results use data that have been reprocessed with this updated lower level calibration.

\section{Summary}
A novel method of performing a PMT timing calibration in large scintillator detectors has been demonstrated through the application of unsupervised learning to the analysis of naturally occurring radioactive backgrounds.
The approach leverages the machinery of deep unsupervised learning to simultaneously and reliably fit over 22,000 calibration parameters in the SNO+ detector.
Within the assumption of a specific PMT timing model, we achieved a timing calibration accuracy of \qty{0.14}{\nano\second} FWHM, based on MC simulation tests. 
Application to physics data from the SNO+ detector was found to reproduce the expected electronic structure of the PMT timing constants.
Using an established SNO+ \ce{^{214}Bi} - \ce{^214Po} coincidence event analysis as a benchmark, the calibration was shown to provide a slight improvement in position resolution compared to the standard SNO+ laserball-based calibration.
It was further able to correctly diagnose changes in calibration parameters that had previously been overlooked. 
This method has thus been shown to provide an accurate and reliable method to perform PMT timing calibrations using only physics data without the resource intensive process of a dedicated hardware-based calibration.

Despite the many physical simplifications including the straight line of flight and the simplified time residual model, the method still achieves a high level of accuracy.
However, if more accuracy is desired or a detector with more complex geometry than SNO+ is used, some more sophisticated modelling may be required.
Due to more complex position dependent optical effects, the time residual model used may need to be more sophisticated and perhaps vary from PMT to PMT. 
The time of flight calculations may also need to account for optical effects such as dispersion and refraction.
Finally, other PMT timing parameters could be calibrated for such as transit time spread by convolving the time residual fit with floated width Gaussians.

The basic approach demonstrated here should not only prove useful for other large scale liquid detectors, but ought to be easily extendable for a wider range of other applications.

\section*{Acknowledgments}
This work was supported by the Science Technology and Facilities Council of
the United Kingdom.
The authors would like to thank the SNO+ collaboration for use of their data, RAT simulation and many useful discussions.

Appropriate representations of the data relevant to the conclusions have been provided within this paper.

\FloatBarrier

\bibliography{biblio}

\end{document}